\documentclass[conference,compsoc]{IEEEtran}
\usepackage{comment}

\ifCLASSOPTIONcompsoc
  \usepackage[nocompress]{cite}
\else
  \usepackage{cite}
\fi
\ifCLASSINFOpdf
  \usepackage[pdftex]{graphicx}
  \graphicspath{{../pdf/}{../jpeg/}}
  \DeclareGraphicsExtensions{.pdf,.jpeg,.png}
\else
  \usepackage[dvips]{graphicx}
  \graphicspath{{../eps/}}
  \DeclareGraphicsExtensions{.eps}
\fi

\usepackage{graphicx}
\usepackage{amsmath}
\usepackage{amssymb}

\ifCLASSOPTIONcompsoc
 \usepackage[caption=false,font=footnotesize,labelfont=sf,textfont=sf]{subfig}
\else
 \usepackage[caption=false,font=footnotesize]{subfig}
\fi

\usepackage{url}

\begin{document}

\title{Fair Influence Maximization in Hypergraphs}

\author{
\IEEEauthorblockN{
Zoë Abhelakh,
Tianrui Mao
and Huijuan Wang*
}

\IEEEauthorblockA{
Faculty of Electrical Engineering, Mathematics, and Computer Science\\
Delft University of Technology\\
Delft, The Netherlands\\
*Email: H.Wang@tudelft.nl
}
}

\maketitle

\begin{abstract}
The Influence maximization problem aims to select a set of seed nodes that maximize the influence, i.e., the average number of influenced nodes, at the end of a spreading process. It has been widely studied with applications in viral marketing, public health campaigns, and social influence. In networks with pronounced community structure, existing approaches often yield an uneven distribution of influenced nodes across communities, which is unfair. Although the fair influence maximization (FIM) problem has been studied for pairwise networks, it remains largely unexplored for hypergraphs, which more accurately represent real-world systems involving group interactions.
We introduce FIMH, a heuristic seed-selection algorithm for FIM on hypergraphs, under the Susceptible–Infected Contact Process (SICP) spreading model. FIMH iteratively estimates the contribution of each candidate node to influence and fairness and selects the node that best trades off these two objectives as an additional seed using a parameter-free utopia-distance criterion. Experiments on seven real-world hypergraphs demonstrate that FIMH achieves an influence comparable to that of state-of-the-art IM methods while significantly reducing influence disparity. Analysis of the topological properties of the selected seed nodes and their contributions to influence and fairness further supports the effectiveness of FIMH.

\end{abstract}

\IEEEpeerreviewmaketitle

\section{Introduction}
  Disease transmission and information diffusion on human contact networks or online social networks can be modelled as spreading processes on networks. In a spreading process, nodes are usually in either an active state (infected, informed, or influenced) or an inactive state (susceptible or unaware). The process begins when an initial set of seed nodes becomes activated at time $t=0$, while all other nodes are inactive. A diffusion model describes the mechanism how each active node activates its neighbours 
\cite{Kempe2003MaximizingNetwork,Li2013EpidemicNetworks,Pastor-Satorras2015EpidemicNetworks,Qu2015SISRates}
which has been widely used to study phenomena ranging from epidemic outbreaks \cite{Yao2022ModelingMaximization}, viral marketing campaigns \cite{Mehmood2016SpheresMarketing}, and the propagation of public health messaging \cite{Yadav2018BridgingYouth}.
Kempe et al. first formalized the Influence Maximization (IM) problem ~\cite{Kempe2003MaximizingNetwork}: given a network, a diffusion model, and a budget $K$, select a set of $K$ seed nodes that maximises the number of activated nodes after T steps. The average number $\sigma(S)$ of activated nodes at step T, when starting with seed set $S$, is referred to as the influence of $S$. It captures how effectively information, behaviour, or epidemics can propagate from a seed set through the network. 

Traditional IM research has focused almost exclusively on pairwise networks, where each link connects exactly two nodes. Kempe et al. ~\cite{Kempe2003MaximizingNetwork} showed that IM problem under the Independent Cascade (IC) or Linear Threshold diffusion models is NP-hard. They introduced the greedy algorithm that starts with an empty seed set and iteratively adding a node that yields the largest marginal gain in influence, $\sigma(S \; \cup \; \{u\}) -\sigma(S)$. The marginal gains are estimated via repeated Monte Carlo simulations of the diffusion process.
The high computational cost of the greedy solution has motivated extensive work to reduce redundant marginal gain computation \cite{Goyal2011CELF++:Networks}. Instead of using simulations to estimate the marginal gain, nodal topological properties (centrality metrics) and analytical approaches are employed \cite{Li2023InfluenceSurvey}.

Many real-world systems cannot be adequately captured by pairwise relationships. For example,  multiple authors can contribute to a single publication, face-to-face interactions, and social media group conversations can involve a group of individuals. These group interactions can be naturally represented using hypergraphs, where nodes represent entities and hyperedges capture the relationship of two or more nodes simultaneously \cite{Bick2021WHAT,Lotito2024MultiplexNetworks, Ceria2022}. Early work on hypergraph IM relied on mapping hypergraphs back to pairwise graphs, losing intrinsic structural information of hypergraphs \cite{Antelmi2021SocialHypergraphs}. Zhu et al.\ \cite{ZhuJianming2019SocialNetworks} proved that IM remains NP-hard on hypergraphs under an IC-style diffusion model.
However, IC diffusion assumes that a newly activated node attempts to activate nodes in all incident hyperedges only at the next time step after its activation.
Suo et al. \cite{Suo2018InformationHypernetworks} introduced the Susceptible–Infected Contact Process (SICP). Specifically, at each time step, each infected node randomly selects one of the hyperedges to which it belongs and infects each susceptible node contained in that hyperedge with infection probability $\beta$. The process starts from an initial seed set $S$ at $t=0$ until time $T$. For SICP on hypergraphs, it has been proved that the influence is monotonous and submodular \cite{Wu2026}. Several IM methods have been developed for SICP. HADP \cite{Xie2023AnHypergraphs} is an adaptive degree-based heuristic in which nodes are selected iteratively according to an adaptive degree, accounting for the number of neighbours that belong to the current seed set. MIE \cite{Gong2024InfluenceEstimation} estimates node influence assuming that infection occurs via paths with few hops and the independence of paths, etc. 

An additional complexity in real-world networks is the community structure. Nodes tend to cluster into communities with dense connections within a community and sparse cross-community connections \cite{Contreras-Aso2023DetectingGraphs, Ruggeri2023CommunityHypergraphs}.
Seed-selection strategies that optimize the global influence may inadvertently induce unequal activation fractions across communities, leaving certain groups systematically under-represented in the activated population. This is especially problematic in the context of public announcements, health campaigns, and critical information, where information should be fairly accessible to all groups. This has motivated the study of fair influence maximization (FIM), which aims to balance the fraction of activated nodes between predefined groups while preserving a high influence \cite{Farnad2020AMaximization, Tsang2019Group-FairnessMaximization, Neophytou2024PromotingSpread}. Existing work on FIM has focused almost exclusively on pairwise networks and fairness notions defined for attribute groups where groups with different attributes could overlap in nodes.

To address this gap, we study FIM on hypergraphs under the SICP spreading model. Our objective is to select a seed set that maximizes the global influence while simultaneously promoting a similar fraction of activated nodes across non-overlapping topological communities. We propose the method \emph{Fair Influence Maximization in Hypergraphs} (FIMH), a scalable, heuristic seed-selection algorithm. At each iteration, FIMH estimates every candidate node's contribution to the influence and the distribution of activated nodes across communities. The algorithm then selects the node that best trades off global influence gain and influence fairness, and adds it to the seed set in a greedy, iterative manner. We evaluate the performance of FIMH, variants of FIMH and state-of-the-art IM baseline methods. Across seven real-world hypergraphs, FIMH achieves influence comparable with best performing IM baseline while significantly improving fair distribution of activated nodes. We further explain how the design of FIMH enables fairer diffusion on hypergraphs without sacrificing overall influence.

Xie et al. \cite{Xie2024FairHypergraphs} recently addressed FIM on hypergraphs.
It differs from our problem in two ways. It employs an IC-type diffusion model rather than SICP, and it targets at fairness across attribute groups that could overlap instead of fairness across non-overlapping topological communities.

\section{Problem Definition}

\subsection{Hypergraphs}
Let the hypergraph be $H=(V,E)$, where $V$ denotes the set of $|V|$ nodes, $E$ the set of $|E|$ hyperedges, respectively. A hyperedge $e_i =(v_1, ..., v_d )$ can connect an arbitrary number $d\ge 2$ of nodes. We consider non-overlapping communities of a hypergraph, represented as $\mathcal{C} = \{C_1,C_2,\dots,C_m\}$ where $m$ denotes the number of communities. Each community $C_i$ contains a subset of nodes, and communities are non-overlapping i.e., $C_i \cap C_{j} = \emptyset$ for $i \neq j$ and $\bigcup_{i=1}^m C_i=V$. A commonly used principle for detecting such communities is modularity maximization. Modularity measures the extent to which nodes within the same community participate in hyperedges together more frequently than would be expected under a random null model that preserves basic structural properties of the hypergraph, such as node degrees and hyperedge sizes. Intuitively, a high modularity indicates that nodes within a community tend to have more interactions with each other than they typically have with the rest of the network. We use the H-louvain algorithm proposed by Kaminski et al \cite{Kaminski2024ModularityHypergraphs} and the public implementation\footnote{\url{https://github.com/pawelwm/h-louvain}} with the settings recommended by the authors to detect communities of real-world hypergraphs. The modularity of the real-world hypergraphs that we consider ranges from 0.42 to 0.76 as seen in Table \ref{tab:dataset-properties}, indicating the evident community structure of these hypergraphs.

\subsection{Spreading process on hypergraphs}\label{sec:diffusion-model}
The influence maximization problem aims to select a set of seed nodes that maximises the influence under a spreading mechanism. We propose to use the Susceptible-Infected (SI) spreading model with Contact Process dynamics in hypergraphs (SICP) \cite{Suo2018InformationHypernetworks} as implemented by Xie et al \cite{Xie2023AnHypergraphs}. Each node is in either susceptible or infected state at any time. Initially, at $t=0$ all nodes in the seed set $S$ are in infected state whereas all the other nodes are susceptible. At each time step $t>0$, each infected node randomly selects one hyperedge it belongs to and infects each susceptible node in the hyperedge independently with infection probability $\beta$. The objective is to identify the seed set that enables fair influence maximization at time step $T$ of the spreading process. 
Infected nodes are also called activated nodes.

\subsection{Fair influence maximization problem}\label{sec:problem-definition}
The Fair Influence Maximization (FIM) problem is defined as follows: given a hypergraph $H=(V,E)$ with a non-overlapping community partition $\mathcal{C}=\{C_1,\dots,C_m\}$, the SICP process, and a budget $K\in\mathbb{Z}^+$, the goal is to select a seed set $S\subseteq V$ with $|S|=K$ nodes that simultaneously maximises the influence and minimises influence disparity at community level at time $T$. The influence maximization objective is defined as:
\begin{equation}
\label{eq:spread-objective}
\textbf{Maximize influence: }
 \arg\max_{S\subseteq V,\;|S|=K}\; \sigma(S).
\end{equation}

Furthermore, we will motivate and define the fairness objective function.
A perfectly fair distribution of activated nodes is achieved when the fraction of activated nodes is identical across all communities \cite{Farnad2020AMaximization}. Let $\sigma_i(S)$ denote the expected number of activated nodes in community $C_i$ at $T$. The activation fraction in community $C_i$ is $y_i(S) = \frac{\sigma_i(S)}{|C_i|}$, where $|C_i|$ is the size of community $C_i$. The global fraction of infection is represented as $ \bar y(S) = \frac{\sigma(S)}{|V|}$. A perfectly fair distribution is obtained when $y_i(S)=\bar y(S), \forall i \in \{1,...,m\}$.

We quantify the deviation from this ideal fair state, called influence disparity, as 
\begin{equation}\label{eq:fair-objective}
   J(S) = \frac{1}{|V|}\sum_{i=1}^m |C_i|\, \bigl| y_i(S) - \bar y(S) \bigr|.
\end{equation}

The corresponding fairness objective is
\begin{equation}
\label{eq:equity-obj}
\textbf{Minimize influence disparity: }
\arg\min_{S\subseteq V,\;|S|=K}\; J(S).
\end{equation}

$J(S)=0$ if and only if $y_i(S)= \bar y(S)$ for every community. The weighting by $|C_i|$ accounts for differences in community size such that $J(S)$ corresponds to the expected disparity experienced by a randomly chosen node. This definition of fairness objective is a natural starting point. Community sizes in networks are typically heterogeneous. Therefore, the fairness objective that aims to maximize the activation fraction of the least-activated group, used for fairness IM across attribute groups, is not suitable here.

\section{Fair Influence Maximization algorithm in Hypergraphs (FIMH)}\label{sec:method}
We propose FIMH, a heuristic greedy algorithm that selects a seed set $S$ aiming to maximise influence under the SICP diffusion model while promoting a fair distribution of activated nodes across communities. 
FIMH derives approximations of the probability $r_j(S\cup\{u\})$ that each node $j$ is infected when any extra node $u$ is added to the seed set $S$. Furthermore, FIMH estimates every candidate node’s contribution to the influence $\sigma(S\cup\{u\}) := \sum_{j \in V} r_j(S\cup\{u\})$ and the distribution of activated nodes across communities. The algorithm then selects the node that best trades off global influence and influence fairness, and adds it to the seed set in a greedy, iterative manner.  

\subsection{Approximation of $r_j(S\cup\{u\})$}
Our approximation assumes that an infected node could only infect its direct neighbours, and that the probability that an infected neighbour subsequently infecting further its own neighbours is negligibly small. This is actually the exact case when $\beta \to 0$.
 In other words, we approximate the probability $r_j(S\cup\{u\})$ given any infection probability $\beta$ by that when $\beta \to 0$. Hence, our approximation is supposed to perform well when $\beta$ is small, as in many real-world applications. 

When $S = \emptyset$ and $r_j(S\cup\{u\})=r_j(u)$, the probability that node $j$ is infected when node $u$ is the single seed node. Furthermore, $r_j(u)=P_{uj}\beta T$, where $P_{uj}$ is called the contact probability and contact refers to the event that a seed node $u$ selects a hyperedge containing $j$. It can be derived \cite{Gong2024InfluenceEstimation} as $P_{uj}= 
\frac{|E(u) \cap E(j)|}{|E(u)|}$, where $E(u)$ denotes the set of hyperedges incident to u.

\begin{figure}[!t]
    \centering

    \subfloat[]{
        \includegraphics[width=0.35\columnwidth]{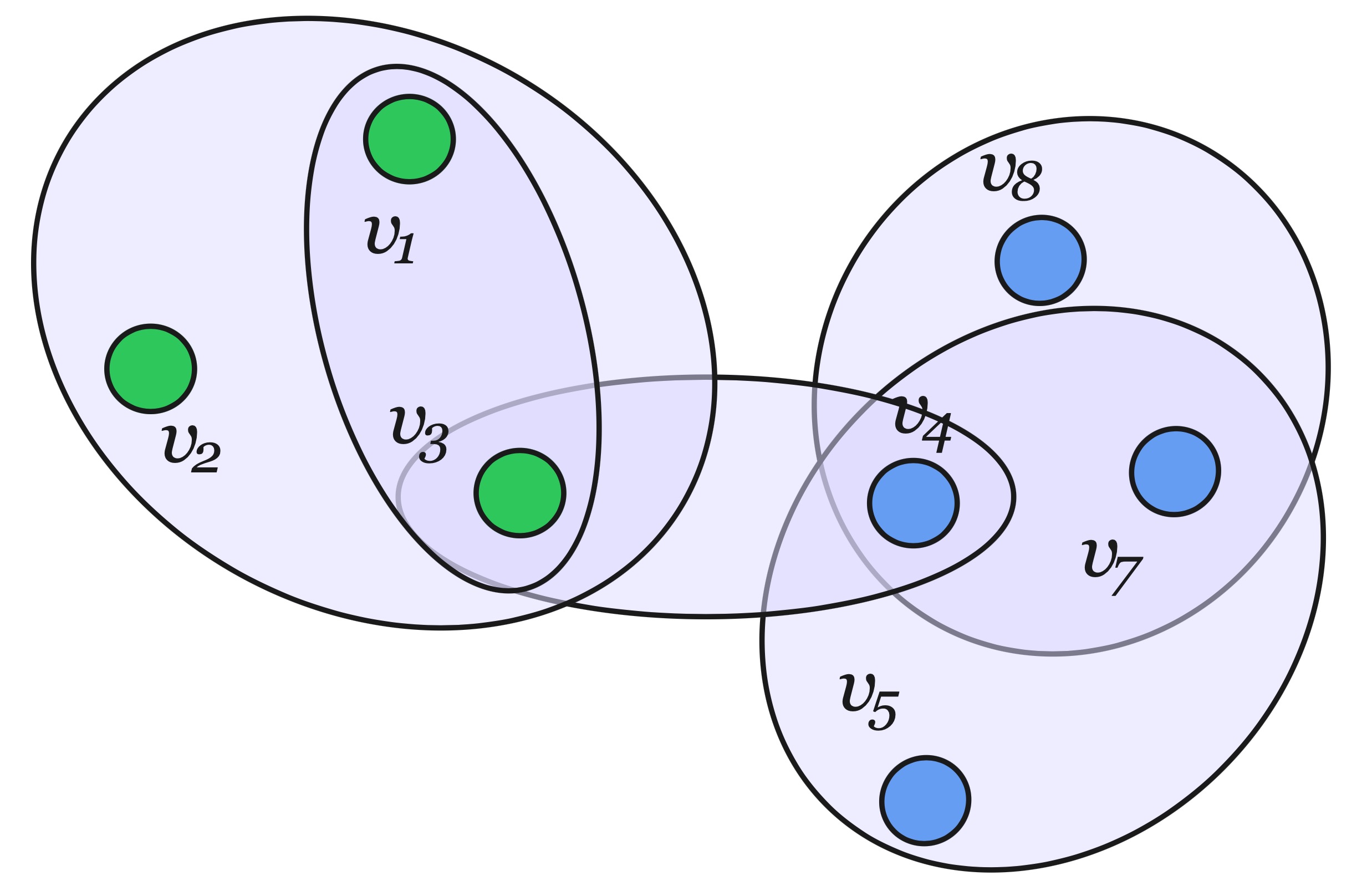}
        \label{fig:hypergraph}
    }
    \hfil
    \subfloat[]{
        \includegraphics[width=0.6\columnwidth]{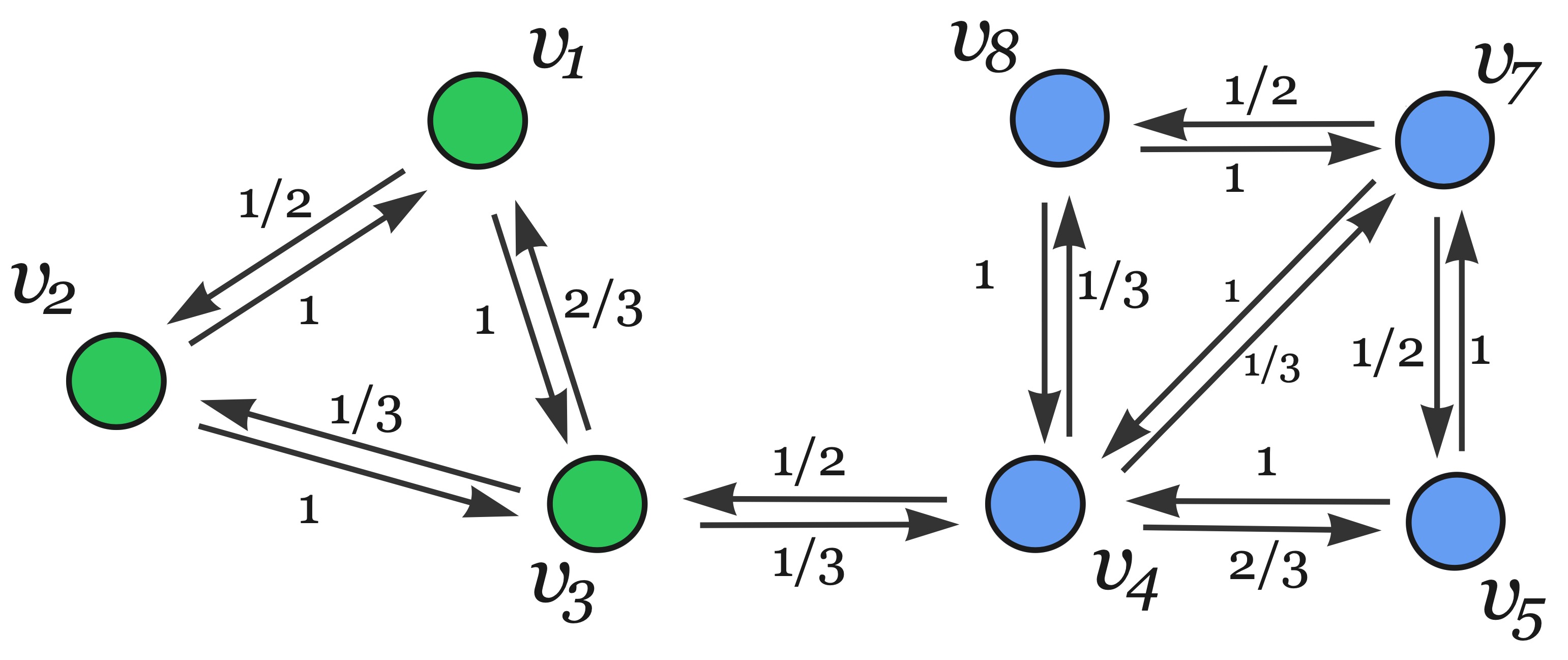}
        \label{fig:dwgraph}
    }
    \caption{(a) Hypergraph $H$ with two communities, in blue and green respectively, and its corresponding contact graph, representing the non-zero elements in the contact probability matrix $P$}
    \label{fig:EPTtransform}
\end{figure}

Figure \ref{fig:EPTtransform} shows a hypergraph as an example and its corresponding contact graph, representing none-zero elements contact probability matrix $P$.

When $\beta \to 0$, the infection processes from any two seed nodes to a target node $j$ are independent. Hence, when $S \neq \emptyset$,  

\begin{align}
\label{eq:state-update}
r_j(S\cup\{u\}) &=
1 - (1 - r_j(S))(1 - P_{uj}\beta T) \\
r_u(S\cup\{u\}) &= 1.
\end{align}
Furthermore, we introduce another assumption in our approximation: $\beta T=1$. Correspondingly, 
\begin{align}
\label{eq:state-update2}
 \hat{r}_j(u)=P_{uj}\\
\hat{r}_j(S\cup\{u\}) &=
1 - (1 - \hat{r}_j(S))(1 - P_{uj}) \\
\hat{r}_u(S\cup\{u\}) &= 1.
\end{align}
FIMH approximates $r_j(S\cup\{u\})$ by $\hat{r}_j(S\cup\{u\})$, the probability that node $j$ belongs to at least one hyperlink of those selected by the seed set $S\cup\{u\}$ in one spreading step, i.e.,  the probability that this node is contacted by the seed set. This approximation becomes exact when $\beta \to 0$ and $\beta T=1$. 

When $\beta$ is large, a seed node can infect a node that is more than one hop away, beyond its neighbors. Hence, we consider the possibility that each seed node could possibly infect nodes within two hops. A variant of FIMH, FIMH (2-hop) approximates $r_j(S\cup\{u\})$ by the probability $\tilde{r}_j(S\cup\{u\})$ that the node can be reached by the seed set within two hops, which is derived in the same way as $\hat{r}_j(S\cup\{u\})$ but replacing the contact matrix $P$ by $\Pi=P+P^2$.

When computing $\tilde{r}_j(S\cup\{u\})$, we have ignored the fact that the paths between $u$ and $j$ and between the seed set $S$ and $j$ within two hops may overlap.  

\subsection{Iterative seed selection}\label{sec:scoring}
FIMH selects seed node iteratively in a greedy way until $K$ seed nodes have been identified. 
At each iteration,  each candidate $u \notin S$ is evaluated according to two criteria: estimated influence $\hat\sigma(S \cup \{u\})=\sum_{j \in V} \hat{r}_j(S\cup\{u\})$ and estimated influence disparity $\hat{J}(S \cup \{u\})$ across communities, which will be derived based on $\hat{r}_j(S\cup\{u\})$.

The estimated influence disparity, the second objective function is defined as:
\begin{equation} \label{eq: disparity_o}
    \hat{J}(S\cup\{u\}) = \frac{1}{|V|} \sum_{i=1}^{m} |C_i|\,\Bigl| \hat{y}_i(S \cup \{u\}) -\hat{\bar y}(S) \Bigr|
\end{equation}
where the estimated global fraction of infection is $\hat{\bar y}(S)=\frac{1}{|V|}\sum_{j\in V} \hat{r}_j(S)$ and the estimated fraction of infection in community $i$ follows $\hat{y}_i(S)=\frac{1}{|C_i|}\sum_{j\in C_i} \hat{r}_j(S)$. This objective function differs slightly from the influence disparity, defined in Eq. \eqref{eq:fair-objective}, which explores the deviation from the updated global average activation fraction $\bar y(S \cup \{u\})$. Our algorithm examines, instead, the deviation from the current average activation fraction $\hat{\bar y}(S)$. The motivatino is that by the objective $\hat{J}(S\cup\{u\})$ tends to be small (large) when $u$ contributes to the activation of nodes in communities whose activation fraction lies below (above) the current activation fraction $\hat{\bar y}(S)$. Later, FIMH (f), another variant of FIMH, considering the objective function \begin{equation}\label{eq:disparity}
    \hat{J}(S\cup\{u\}) = \frac{1}{|V|} \sum_{i=1}^{m} |C_i|\,\Bigl| \hat{y}_i(S \cup \{u\}) -\hat{\bar y}(S\cup \{u\}) \Bigr|
\end{equation}
will be evaluated later.

At each iteration, each candidate node $u \notin S$ is represented by the pair $(\hat\sigma(S\cup\{u\}),\hat{J}(S\cup\{u\}))$ denoting its estimated influence and influence disparity. FIMH
applies per-iteration min-max normalisation to each objective among all candidate nodes, respectively, mapping each candidate to $(\sigma^*(u),J^*(u))$, where $\sigma^*(u) \in [0,1]$ and $J^*(u) \in [0,1]$. A larger $\sigma^*(u)$ indicates a relatively high influence induced by candidate node $u$ among all candidates at that iteration. This normalization is motivated by the fact that the influence increases with the size of the seed set, whereas influence disparity does not.

To guide the multi-objective decision, we adopt the classical notion of an ideal (or utopia) point from multi-objective optimization \cite{Miettinen1998NonlinearOptimization}. The utopia point combines, for each objective, the best achievable value and serves as an infeasible benchmark representing the ideal trade-off. In normalized coordinates, this reference point is $(1,0)$, corresponding to maximal influence and minimal influence disparity at a given iteration. FIMH chooses as the next seed the candidate $u^\star$ with smallest Euclidean distance to the utopia point:
\begin{equation}\label{eq:utopia}
    u^\star = \arg \min_{u \notin S}\sqrt{(1-\sigma^*(u))^2+J^*(u)^2}
\end{equation}

\subsection{Lifting the fairness objective at the first iteration}
When $S=\emptyset$, all community activation fractions are zero and the influence disparity measure $\hat{J}(S)=0$, corresponding to a state of perfect fairness but zero influence. In this setting, the fairness objective prefers a seed node that leads to minimal influence, against the objective of maximizing the influence. Hence, we lift the fairness objective in the first iteration and select the initial seed that leads to the highest estimated influence. From the second iteration onward, the utopia-distance rule is applied, allowing the algorithm to explicitly trade off additional influence and disparity.

\subsection{Variants of FIMH}
We consider as well three variants of FIMH. The first one, FIMH(f), as defined earlier, considers a slightly differenct fairness disparity objective.
The second variant FIMH(l), differs from FIMH in the multi-objective decision. FIMH(l) chooses the candidate $u^*$ that minimizes the linear function:
\begin{equation}\label{eq:linear}
u^\star=\arg\min_{u}\Big(\alpha\,J^*(u)-(1-\alpha)\sigma^*(u)\Big),
\qquad \alpha\in[0,1].
\end{equation}
We consider $\alpha=0.5$ for FIMH(l), reflecting a neutral trade-off between the two objectives.

The third variant FIMH(2-hop), as defined before, approximates the infection probability of a node by a seed set by the probability that the node can be reached by the seed set within two hops, instead of the one-hop reachability as in FIMH.

\subsection{Complexity analysis}
The time complexity of the FIMH algorithm is primarily determined by the computation of the contact probability matrix $P$ and by the iterative greedy process used to select seed nodes. Recall that $|V|$ and $|E|$ denote the number of nodes and hyperedges in the hypergraph respectively and $K$ represents the seed budget, i.e., the size of the target seed set. Computing the contact probability matrix $P\in [0,1]^{|V| \times |V|}$ requires $O(|V|^2)$ time. At each iteration of the greedy algorithm, evaluating the marginal influence gain and influence disparity for all candidates takes $O(|V|)$ time, because we update the impact of adding each candidate $u$ on the objectives using $P$. Thus, the time complexity per iteration is $O(|V|^2)$. Since the algorithm runs for $K$ iterations, the overall time complexity of FIMH is $O(K |V|^2)$.

\section{Performance Evaluation Method} 
We evaluate the performance of the proposed algorithm FIMH, its variants and established baselines  in seven real-world hypergraphs in terms of both influence and fairness.

\subsection{Datasets}
Seven real-world hypergraph datasets\footnote{\url{https://www.cs.cornell.edu/~arb/data/}} are considered, widely used in prior hypergraph analysis and influence maximization studies \cite{Xie2023AnHypergraphs}\cite{Gong2024InfluenceEstimation}\cite{FrancescoLotito2024HyperlinkNetworks}. The topological properties of the datasets are shown in Table \ref{tab:dataset-properties}. A brief description of each dataset is given. \textbf{Algebra \& Geometry \cite{Veldt2020MinimizingHypergraphs}} are two hypergraphs derived from MathOverflow users answering questions. Hyperedges correspond to groups of users responding to the same question. \textbf{Restaurant-Rev \& Bars-Rev \cite{Amburg2020FairGroups}}: Respectively, restaurants and bars are divided into categories. Yelp users who review the same restaurant/bar category in a month form a hyperedge. \textbf{Music-Rev} \cite{Ni2019JustifyingAspects}: Amazon users reviewing the same blues-music category within a month form a hyperedge.
 \textbf{Contact-high-school \& Contact-primary-school} \cite{Stehle2011High-resolutionSchool, Chodrow2021GenerativeModularity} are derived from wearable proximity sensors in French schools. Each hyperedge represents a group of students simultaneously in close proximity. 

\begin{table*}[!t]
\caption{Topological properties of the datasets. $n$ and $m$ represent the number of nodes and hyperedges in the hypergraph. $\langle deg\rangle$ is the average node degree, $\langle d^H\rangle$ is the average node hyperdegree, $\langle d^E\rangle$ represents the number of nodes in a hyperedge on average. $C$ denotes the number of detected communities, with $C_{\min}$ and $C_{\max}$ the smallest and largest community sizes, and $\langle C\rangle$ the average community size. $q_H$ denotes the hypergraph modularity}
\label{tab:dataset-properties}

\centering

\begin{tabular}{lrrrrrrrrrr}
\hline
\bfseries Dataset & \bfseries $n$ & \bfseries $m$ & \bfseries $\langle deg\rangle$ & \bfseries $\langle d^H\rangle$ & \bfseries $\langle d^E\rangle$ & \bfseries $C$ & \bfseries $C_{min}$ & \bfseries $C_{max}$ & \bfseries $\langle C\rangle$ & \bfseries $q_H$ \\
\hline
Algebra & 423 & 1268 & 78.90 & 19.53 & 6.52 & 10 & 2 & 88 & 42.30 & 0.42 \\
Geometry & 580 & 1193 & 164.79 & 21.53 & 10.47 & 21 & 1 & 137 & 27.62 & 0.49\\
Music-Rev & 1106 & 694 & 167.88 & 9.49 & 15.13 & 24 & 2 & 361 & 46.08 & 0.70 \\
Restaurants-Rev & 565 & 601 & 79.75 & 8.14 & 7.66 & 7 & 11 & 177 & 80.71 & 0.76 \\
Bars-Rev & 1234 & 1194 & 174.30 & 9.62 & 9.94 & 37 & 4 & 314 & 33.35 & 0.59\\
High-school & 327 & 7818 & 35.58 & 55.63 & 2.33 & 8 & 33 & 67 & 40.88 & 0.63\\
Primary-school & 242 & 12704 & 68.74 & 126.98 & 2.42 & 6 & 24 & 51 & 40.33 & 0.42\\
\hline
\end{tabular}

\end{table*}

\subsection{Baselines}
To assess the performance of our approach, we compare it against several classic and state-of-the-art baselines. \textbf{Greedy.} The greedy influence maximization algorithm \cite{Kempe2003MaximizingNetwork} starts from an empty seed set and iteratively adds a node that yields the largest marginal gain in expected influence, which is estimated via repeated Monte Carlo simulations of the diffusion process. We perform $500$ Monte Carlo runs per iteration and report the results only if the total runtime is within 48 hours on the Delft AI Cluster. \textbf{HADP.} The Hyper-Adaptive Degree Pruning method \cite{Xie2023AnHypergraphs} iteratively selects nodes according to an adaptive degree score that discounts nodes whose neighbours are already in the neighbourhood of nodes in the current seed set. This reduces redundancy by discouraging the selection of nodes with highly overlapping influence regions.

\textbf{MIE.} Multi-hop Influence Estimation \cite{Gong2024InfluenceEstimation} estimates the influence of a node using a probability model to approximate the likelihood of a node infecting another node under SICP diffusion process. Seed nodes are selected greedily according to the resulting estimated marginal influence gain. Just like HADP, MIE also accounts for nodes that are likely to already be influenced by previously selected seeds. \textbf{Adeff}. Adaptive Neighbourhood Coefficient algorithm \cite{Gong2024InfluenceEstimation} ranks nodes by the average size of hyperlinks incident to a node and choose nodes with the largest average size as the seed set. 
\textbf{Hyperdegree.} It selects nodes with the highest hyperdegree, i.e., the highest number of incident hyperlinks.  \textbf{Degree.} It selects nodes with the highest degree, i.e., the number of distinct nodes connected to a node via hyperlinks.

Following prior work~\cite{Xie2023AnHypergraphs, Gong2024InfluenceEstimation}, we evaluate all algorithms for seed size $K\in [1,25]$ when $(\beta=0.01, \; T=25)$,$(\beta=0.015, \; T=15)$ and $(\beta=0.02, \; T=10)$. Given a set of these parameters, a hypergraph and a seed set, influence and influence disparity are derived as the average of 250 independent realizations of the SICP process.

\section{Results}
We evaluate FIMH and its three variants in terms of influence and influence disparity. Afterwards, the best performing algorithm FIMH and FIMH(2-hop) will be further compared with all baselines. Unless stated otherwise, results corresponding to $\beta=0.01$, $T=25$ are used to demonstrate our findings. Results of other parameter sets will be given in Subsection~\ref{FIMHbaseline}, leading to the same observations. 

\subsection{Comparison of FIMH and its variants} 

\begin{figure*}[!t]
    \centering
    \includegraphics[width=\textwidth]{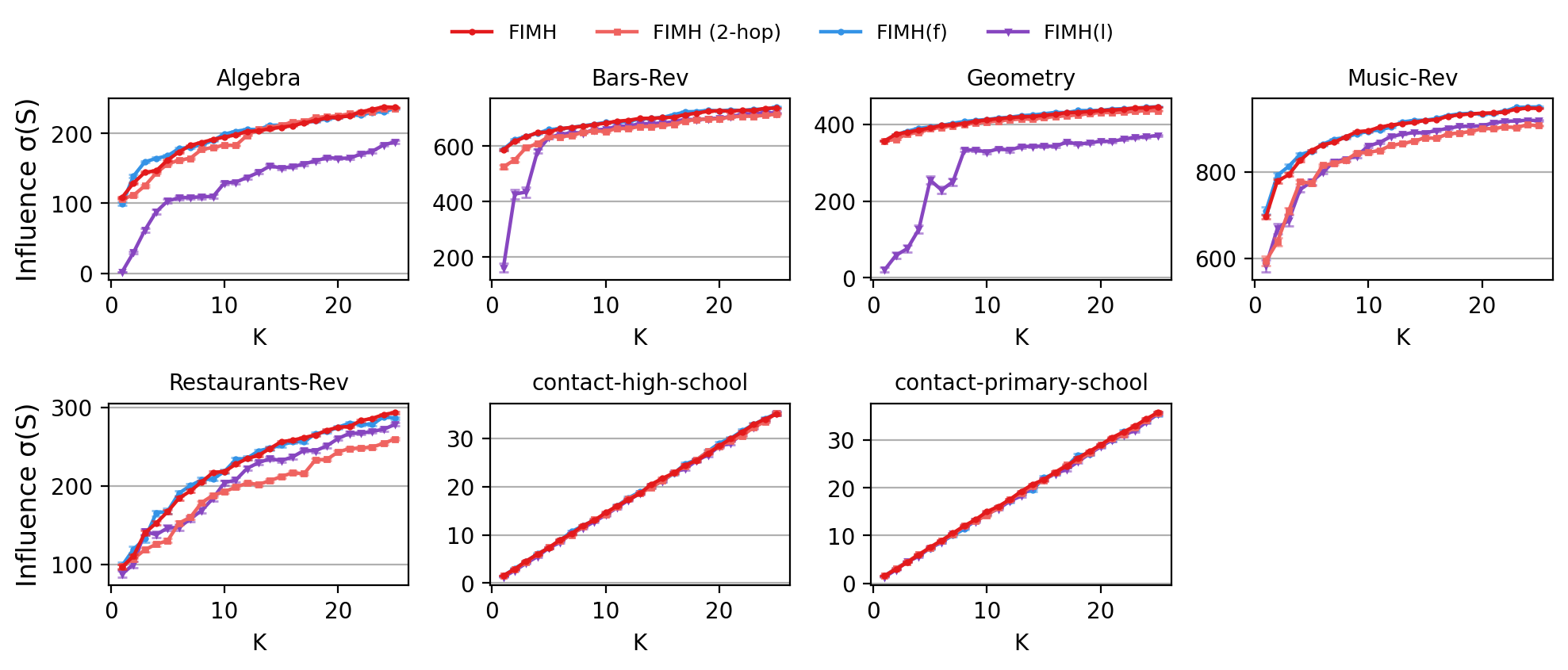}
    \caption{influence $\sigma(S)$ as a function of seed budget $K \in [1,25]$ for FIMH and its variants under SICP with $\beta=0.01$ and $T=25$.}
    \label{fig:spread_Vs_K_variants}
\end{figure*}

\begin{figure*}[!t]
    \centering
    \includegraphics[width=\textwidth]{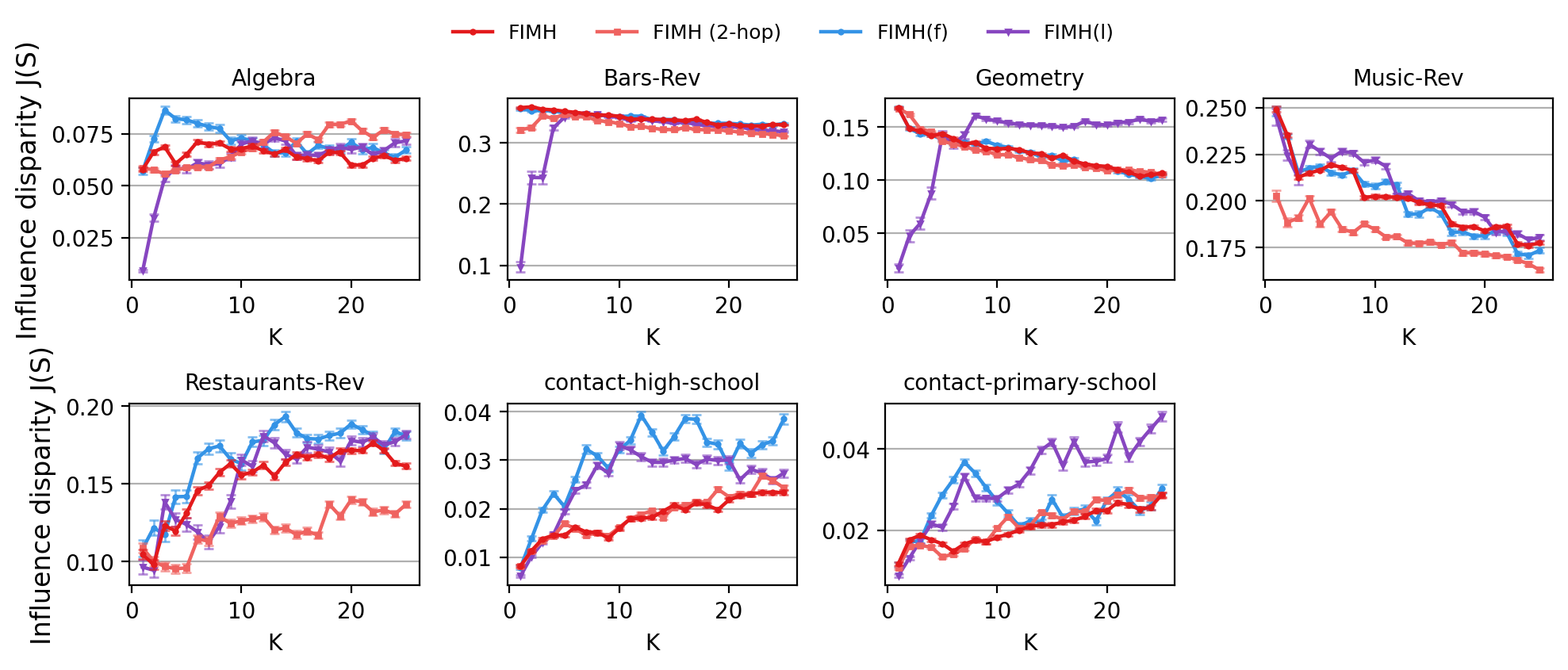}
    \caption{influence disparity $J(S)$ as a function of seed budget $K \in [1,25]$ for FIMH and its variants under SICP with $\beta=0.01$ and $T=25$.}
    \label{fig:equity_vs_K_variants}
\end{figure*}

Figure \ref{fig:spread_Vs_K_variants} and Figure \ref{fig:equity_vs_K_variants} compare FIMH with its three variants, FIMH(f), FIMH(l), and FIMH(2-hop), in terms of influence and influence disparity at time $T$, the end of the spreading process. 
In terms of influence, FIMH and FIMH(f) perform the best and similarly across all datasets whereas FIMH leads to an evidently lower disparity than FIMH(f). Disparity objective function Eq.\eqref{eq: disparity_o} considered by FIMH 
preferentially selects new seed nodes that boost activation in communities whose activation fraction falls below this average. It effectively reduces the influence disparity compared to the objective disparity function Eq.\eqref{eq:disparity} employed by FIMH(f) that
could choose seed nodes that contribute more to over-activated communities.

In terms of fairness, FIMH and FIMH(2-hop) perform the best, whereas FIMH(2-hop) slightly outperforms FIMH. In view of the better performance of FIMH and FIMH(2-hop) in influence and fairness, they will be compared further with the baselines.

In network Contact-high-school and Contact-primary-school, all seed selection methods perform the same in influence. This is because these networks have evidently lower average size of a hyperlink, as shown in Table \ref{tab:dataset-properties}, such that the seed nodes can hardly infect other nodes, when $\beta=0.01$ and $T=25$ are small. However, FIMH and its variants lead to different influence disparities, reflecting mainly the disparity of seed nodes allocated in communities.

\subsection{Comparison of FIMH and baselines}
\label{FIMHbaseline}
\begin{figure*}[!t]
    \centering
    \includegraphics[width=\textwidth]{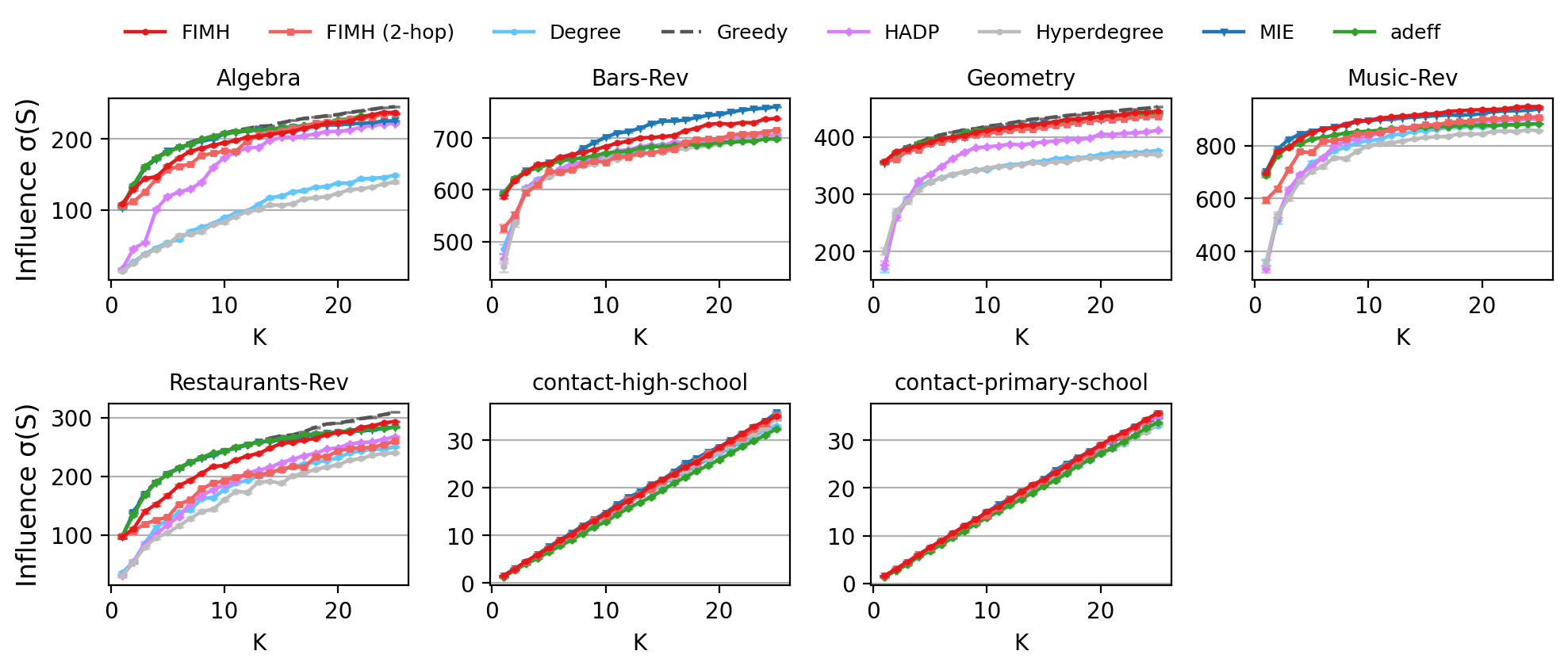}
    \caption{influence $\sigma(S)$ as a function of seed budget $K \in [1,25]$ for all methods under SICP with $\beta=0.01$ and $T=25$.}
    \label{fig:spread_Vs_K}
\end{figure*}

\begin{figure*}[!t]
    \centering
    \includegraphics[width=\textwidth]{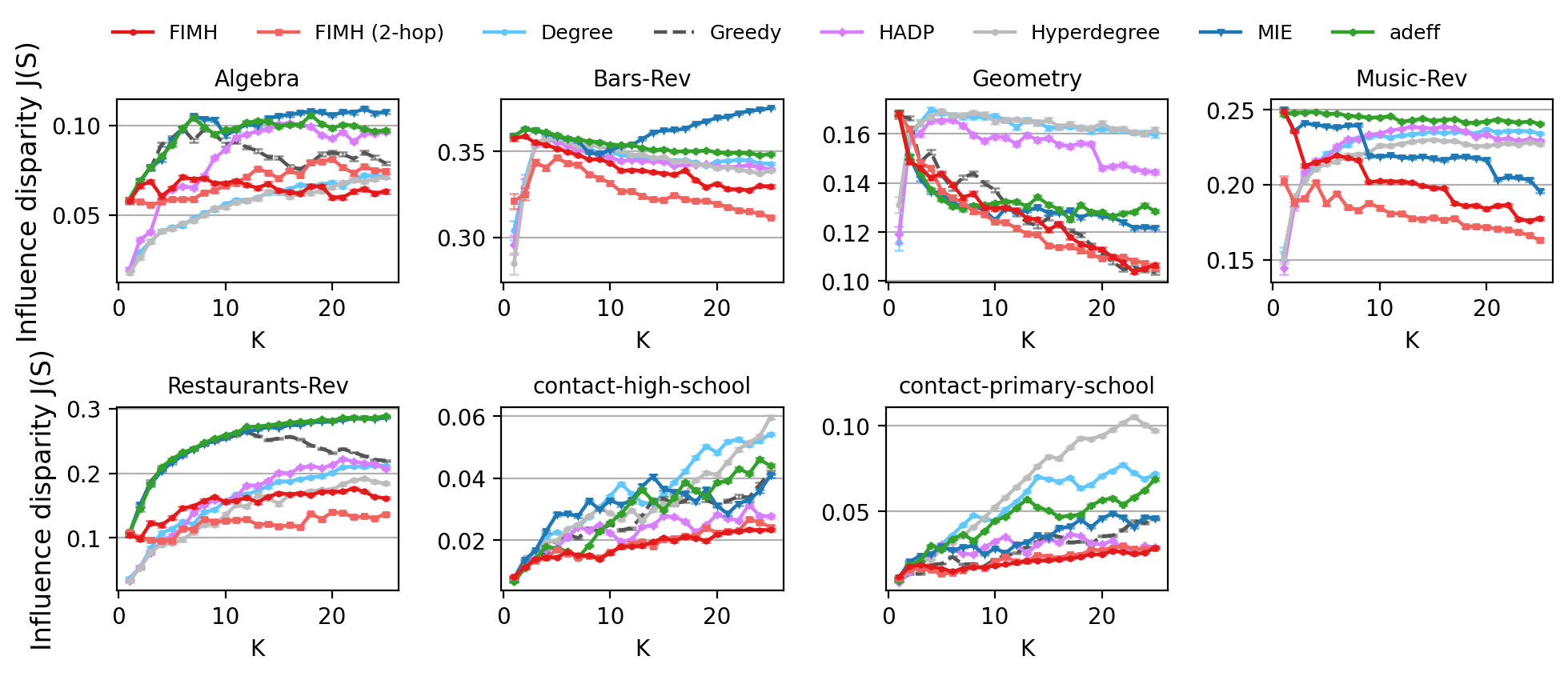}
    \caption{influence disparity $J(S)$ as a function of seed budget $K \in [1,25]$ for all methods under SICP with $\beta=0.01$ and $T=25$.}
    \label{fig:equity_vs_K}
\end{figure*}

Figure \ref{fig:spread_Vs_K} reports the influence $\sigma(S)$ as a function of $K$ for FIMH, FIMH(2-hop) as well as the baselines. To summarize the performance of a method across all $K$, Table \ref{tab:AUC-Spread-K} reports the Area Under the Curve (AUC) score, i.e., the sum of $\sigma(S)$ over $K$. A larger AUC indicates better overall influence across the full range of $K = [1, 25]$.  The performance of the Greedy algorithm is reported if the simulation is completed within 48 hours. The greedy algorithm remains a strong upper bound in terms of influence but performs poorly in fairness, according to Figure \ref{fig:equity_vs_K}. It will not be discussed further nor included in the AUC Tables for influence and disparity respectively, due to its high computational complexity. 

With respect to influence, FIMH and MIE  is among the best performing methods across all datasets. The FIMH (2-hop) variant exhibits slightly lower influence but remains mostly close to the best method. For each network, we identify the highest AUC achieved across all the methods and compute the relative performance of each method as its AUC normalized by the highest AUC. FIMH performs the best or second best in most networks, achieving a relative performance $93\% - 100\%$ of the best observed influence AUC and $98,02\%$ on average. FIMH (2-hop) reaches a relative performance $93.49\%$ on average, ranging within $80.85\% - 98.53\%$. These results show that incorporating fairness into the seed selection does not evidently reduce the influence. 

Several baselines (HADP, MIE, Adeff) incorporate preliminary mechanisms to reduce the overlap of nodes that could be potentially activated by two seed nodes to maximize influence. For example, HADP iteratively selects the node that have the largest number of neighbors that do not belong to the existing seed set. MIE after including a node to the seed set, it will call a function to prune a subset of the nodes, that are close to the selected seed node, so that those will not be picked as seeds in the next iteration. These mechanisms to reduce the overlap of nodes that could be activated by two seed nodes may implicitly contribute to fairness.  In contrast, FIMH and FIMH (2-hop) explicitly favour seed nodes, that improve the activation fractions in under-activated communities.

Figure \ref{fig:equity_vs_K} illustrates the influence disparity $J(S)$ as a function of $K$, while Table \ref{tab:AUC-equity-K} summarises the corresponding AUC scores. 
For influence disparity, FIMH and FIMH (2-hop) perform evidently better than baselines, whereas FIMH (2-hop) achieves on average a lower influence disparity. 

Among the baselines, MIE tends to result in the highest influence. Compared to MIE, FIMH performs comparably in influence but reduces the disparity by $24.3\%$ on average. 

This confirms that explicit seeding in hypergraphs with fairness as an objective can substantially improve fairness without sacrificing influence. The same can be observed, when considering another two sets of parameters that have been used in prior hypergraph IM work \cite{Xie2023AnHypergraphs,Gong2024InfluenceEstimation}: $(\beta = 0.015, T = 15)$ and $(\beta = 0.02, T = 10)$. The corresponding performance in AUC with respect to influence and influence disparity is given in Appendix \ref{app:sensitivity-analysis}. 

\begin{table*}[!t]
\caption{AUC of influence computed for each algorithm curve in Figure \ref{fig:spread_Vs_K} when $\beta=0.01, T=25$. The highest AUC in each dataset is shown in bold and the second-highest is marked with an asterisk (*).}
\label{tab:AUC-Spread-K}

\centering
\resizebox{\textwidth}{!}{
\begin{tabular}{lrrrrrrr}
\hline
\bfseries Dataset & \bfseries FIMH & \bfseries FIMH (2 hop) & \bfseries HADP & \bfseries ADEFF & \bfseries MIE & \bfseries Degree & \bfseries Hyperdegree \\
\hline
Algebra & 4716.40 & 4603.47 & 4027.05 & \textbf{4911.78} & 4839.55* & 2385.64 & 2207.94 \\
Geometry & \textbf{9973.73} & 9826.78 & 8908.33 & 9947.92* & 9925.04 & 8243.39 & 8200.32 \\
Bars-Rev & 16583.82* & 15888.27 & 15925.92 & 16119.99 & \textbf{16942.71} & 15882.93 & 15713.41 \\
Music-Rev & \textbf{21439.88} & 20184.40 & 19534.35 & 20406.76 & 21335.77* & 19200.07 & 18624.66 \\
Restaurants-Rev & 5451.77 & 4693.54 & 4598.62 & \textbf{5805.45} & 5782.08* & 4397.43 & 4056.13 \\
contact-high-school & 448.15* & 444.33 & 437.25 & 405.87 & \textbf{454.92} & 427.45 & 435.10 \\
contact-primary-school & 454.52* & 450.47 & 443.24 & 423.50 & \textbf{455.40} & 427.69 & 428.55 \\
\hline
\end{tabular}
}
\end{table*}

\begin{table*}[!t]
\caption{AUC of influence disparity computed for each algorithm curve in Figure \ref{fig:equity_vs_K} when $\beta=0.01, T=25$. The lowest AUC in each dataset is shown in bold and the second-lowest is marked with an asterisk (*).}
\label{tab:AUC-equity-K}
\centering
\resizebox{\textwidth}{!}{
\begin{tabular}{lrrrrrrr}
\hline
\bfseries Dataset & \bfseries FIMH & \bfseries FIMH (2 hop) & \bfseries HADP & \bfseries ADEFF & \bfseries MIE & \bfseries Degree & \bfseries Hyperdegree \\
\hline
Algebra & 1.57 & 1.66 & 1.98 & 2.29 & 2.37 & 1.36* & \textbf{1.33} \\
Geometry & 3.03* & \textbf{2.97} & 3.73 & 3.18 & 3.13 & 3.91 & 3.93 \\
Bars-Rev & 8.17* & \textbf{7.85} & 8.26 & 8.49 & 8.68 & 8.31 & 8.29 \\
Music-Rev & 4.82* & \textbf{4.32} & 5.45 & 5.86 & 5.33 & 5.45 & 5.30 \\
Restaurants-Rev & 3.71 & \textbf{2.91} & 3.99 & 6.09 & 6.02 & 3.85 & 3.39* \\
contact-high-school & \textbf{0.44} & 0.45* & 0.56 & 0.69 & 0.74 & 0.85 & 0.77 \\
contact-primary-school & \textbf{0.50} & 0.52* & 0.68 & 1.07 & 0.82 & 1.29 & 1.55 \\
\hline
\end{tabular}
}
\end{table*}

\subsection{Performance explanation}\label{sec:influence-fairnessrelationship}
In this subsection, we explore why FIMH achieves a more fair distribution of activated nodes in communities while being among the best performing methods in influence. 
\begin{figure}[!t]
    \centering
    \subfloat[Algebra]{
        \includegraphics[width=\columnwidth]{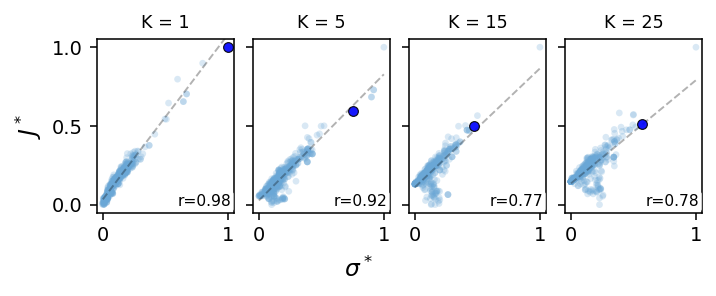}
        \label{fig:snapshot_algebra}
    }
    \vspace{0.5em}
    \subfloat[Contact-primary-school]{
        \includegraphics[width=\columnwidth]{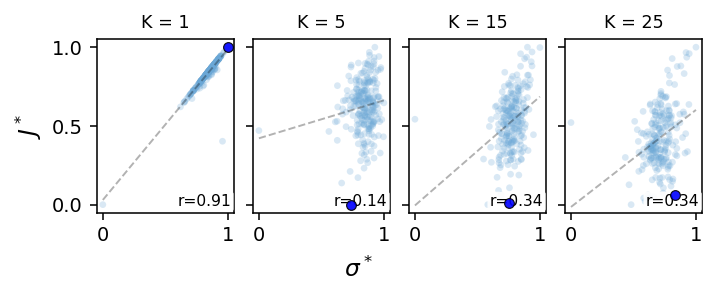}
        \label{fig:snapshot_restaurants-rev}
    }
    \caption{Relative influence $\sigma^*(u)$ versus relative influence disparity $J^*(u)$ of each candidate node at the FIMH selection step $K \in \{1,5,15,25\}$ when $\beta=0.01$, their Pearson correlation $r$ and linear fit in dashed line in network (a) Algebra and (b) contact primary school. The blue dot represents the node closest to the utopia point, thus selected as a seed.}
    \label{fig:spread-equity-two-datasets}
\end{figure}
A key design of FIMH is the multi-objective decision, i.e., the objective function defined by Eq. \eqref{eq:utopia}. We first examine 
how this multi-objective decision balances the two objectives: influence versus fairness. In each iteration, FIMH estimates the relative influence and influence disparity $(\sigma^*(u),J^*(u))$ obtained by each candidate node. It selects as the next seed the candidate with the smallest objective function Eq. \eqref{eq:utopia}, i.e., the candidate that has the smallest Euclidean distance to the utopia point $(1,0)$. Figure \ref{fig:spread-equity-two-datasets} shows the candidate distributions per-iteration in the $(\sigma^*(u),J^*(u))$  plane for two representative datasets (Algebra and Contact-high-school) in the selection rounds $K\in\{1,5,15,25\}$. In all datasets, we observe a strong positive correlation between influence $\sigma^*(u)$ and influence disparity $J^*(u)$ induced by a candiate node in early iterations: nodes leading to a large influence $\sigma^*(u)$ tend to incur a high influence disparity $J^*(u)$. When $K$ is small. The Pearson correlation between $\sigma^*(u)$ and  $J^*(u))$ exceeds 0.9 for all datasets (Appendix \ref{app:spread-equity-trade-off}). By design, FIMH chooses the candidate node with the highest influence $\sigma^*(u)$ as the first seed node, and since $K>2$ it selects the node that minimizes the objective function Eq. \eqref{eq:utopia}. Figure \ref{fig:spread-equity-two-datasets} illustrates that when $K>2$, FIMH possibly chooses a node that has a relatively low influence disparity and low influence as the seed. This shows that FIMH indeed balances the two objectives: influence and influence disparity. 

Furthermore, we explore the network locations of the seed sets selected by different methods. Specifically, two topological properties of the seed nodes selected by each method will be investigated: their mutual topological closeness and their distribution across communities. 

\begin{figure*}[t!]
    \centering
    \includegraphics[width=1\textwidth]{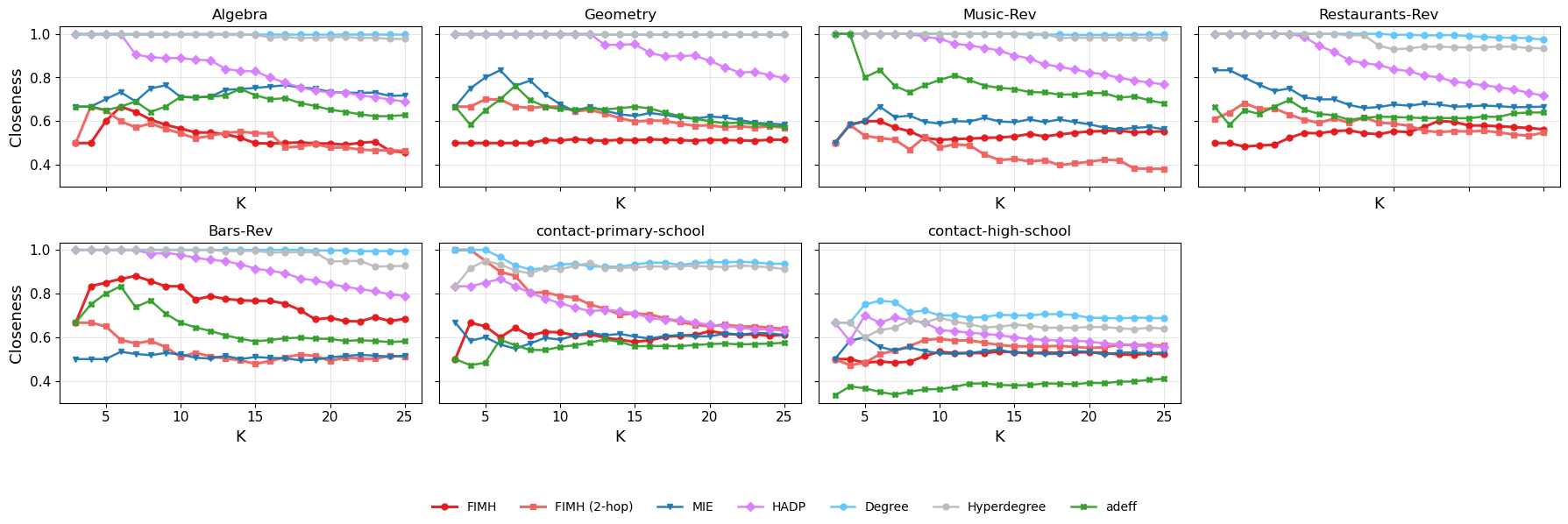}
    \caption{Mutual closeness of a seed set of size $K$ selected by each method in each dataset.}
    \label{fig:closeness}
\end{figure*}

A path between two nodes in a hypergraph is a sequence of nodes, where each pair of consecutive nodes belong to a unique hyperedge \cite{Nortier2025}. In line with the definition of closeness centrality of a node in pairwise networks, the mutual closeness of a selected seed set of size $K$ in a hypergraph is defined as 
$\frac{2}{K(K-1)}\sum_{i\neq j, i,j \in S}\frac{1}{H_{ij}}$, where $H_{ij}$ is the hopcount of the shortest path between $i$ and $j$. The hopcount of a path between two nodes is the number of hyperlinks traversed and the shortest path refers to the path with the smallest hopcount. A large closeness implies that seed nodes are located closely from each other on average in topology. As discussed earlier, baseline methods have incorporated preliminary mechanisms to reduce the overlap of nodes that could potentially be activated by two seed nodes to maximize influence. A seed set with a small closeness could possibly reduce the overlap. Moreover, seed set with a small closeness may also benefit the influence fairness. As shown in Figure \ref{fig:closeness}, seed nodes selected by FIMH and FIMH(2-hop) tend to be further away from each other than the other methods in most networks. 

\begin{figure*}[t!]
    \centering
    \includegraphics[width=1\textwidth]{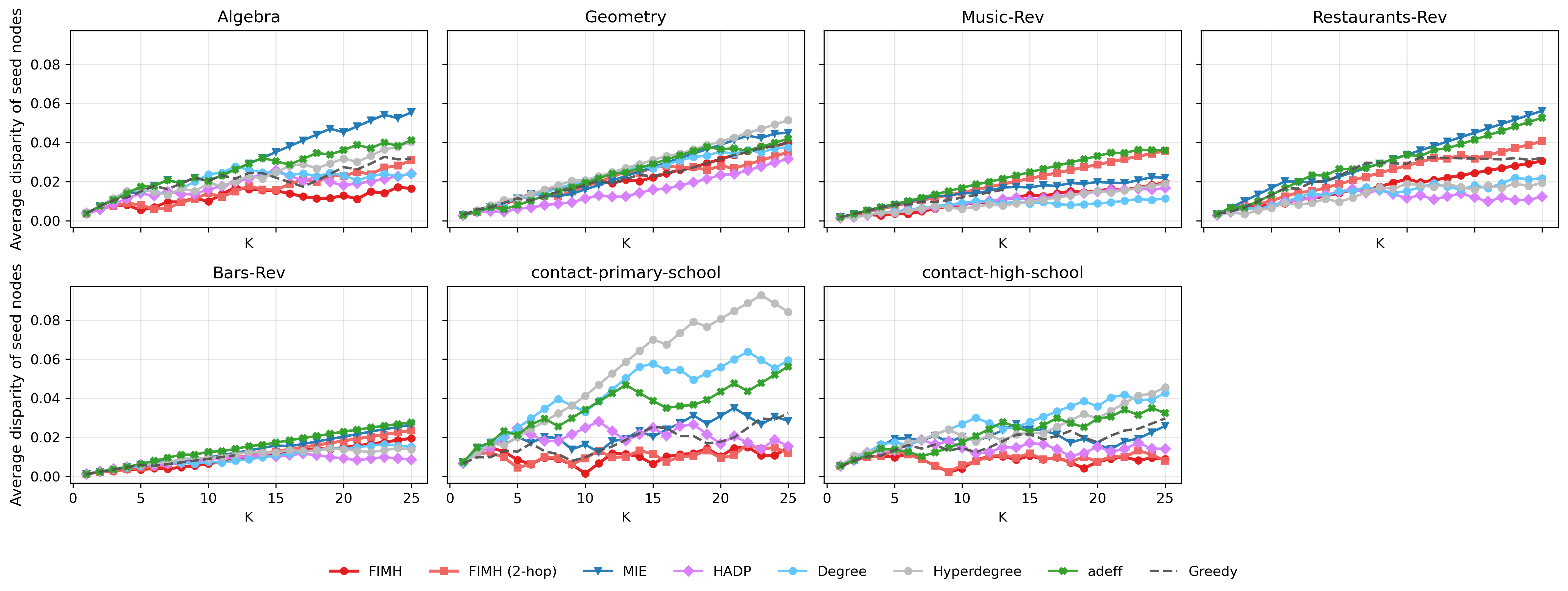}
    \caption{Disparity of seed nodes selected by each method as a function of K in each dataset. }
    \label{fig:disparity_of_seed_set}
\end{figure*}

Is a seed set itself allocated to communities in a fair way? How is the fairness of the seeds themselves related to our target influence fairness, i.e., the fairness of all activated nodes by the end of a spreading process? Figure \ref{fig:disparity_of_seed_set} demonstrates the disparity of the seed set itself, selected by each method in each dataset. The disparity of a seed set is calculated using the same definition as in Eq. \eqref{eq:fair-objective}, but with only the seed nodes considered activated. In network contact-primary-school and contact-high-school, the seed sets selected by FIMH and FIMH(2-hop) have the lowest disparity compared with the other baselines. In these networks, few nodes get infected by a seed set, as shown in Figure \ref{fig:spread_Vs_K}. Hence, the influence disparity is largely determined by the disparity of a seed set. This explains the outperformance of FIMH and FIMH(2-hop) in fairness in these two networks. 

In the other datasets, seed sets chosen by FIMH, FIMH(2-hop) and HADP tend to have the lowest disparity. However, FIMH, FIMH(2-hop) perform the best among all methods in fairness whereas HADP is among the worst performing ones. In these networks, the number of nodes activated is far larger than the seed set. HADP iteratively selects the node with the largest number of neighbours that do not belong to the existing seed set as the new seed node. This contributes to the low disparity of the seed set but without considering the disparity of activated nodes. In contrast, FIMH and FIMH(2-hop) achieve high influence fairness by explicitly incorporating influence disparity into the objective function and by estimating disparities in node activation at each step of the seed selection process.

\section{Conclusion}
In this paper, we explore the problem of fair influence maximization on hypergraphs, where fairness is quantified through the disparity of the fraction of activated nodes at the end of the spreading process across communities. We propose FIMH (and its variants), seed selection method that jointly optimises influence and fairness. FIMH iteratively estimates the influence and influence disparity induced by each candidate node and selects the candidate that best balances these two objectives according to the utopia-distance rule as an additional seed. Experiments on seven real-world hypergraphs demonstrate that FIMH achieves influence comparable with state-of-the-art baselines designed for influence maximization while substantially reducing influence disparity across communities. These results show that, although fairness and influence maximization are generally negatively correlated at the candidate level, enforcing fairness as an additional objective does not substantially compromise the performance in influence maximization. Seeds selected by FIMH tends to be located far away from each other and fairly across communities. While these factors contribute to FIMH's good performance, the design of FIMH in estimating influence and influence disparity and the multi-objective decision is the key.  

Several directions remain open for future work. First, better estimation of the infection probability of any node by a given seed set is needed especially when the infection probability of the SICP process is large, thus spreading trajectory could be more than two hops. Second, our fairness formulation assumes a non-overlapping community partition: extending the framework to overlapping, dynamic, or attribute-based group definitions is a natural next step. Finally, adapting the FIMH framework to other diffusion models beyond SICP offers a promising direction for further research on fair influence maximization in hypergraphs or temporal hypergraphs. 

\ifCLASSOPTIONcompsoc
  \section*{Acknowledgments}
\else
  \section*{Acknowledgment}
\fi

We thank for the support of Netherlands Organisation for Scientific Research NWO (project FORT-PORT no. KICH1.VE03.21.008, project NEPTARGOS no. KICH1.VE05.23.003), NExTWORKx, a collaboration between TUDelft and KPN on future telecommunication networks and Delft AI Cluster at TUDelft.

\newpage

\bibliographystyle{IEEEtran}
\bibliography{references_v2}

\appendix

\section{Influence $\sigma^*(u)$ and influence disparity $J^*(u)$ of a candidate node}\label{app:spread-equity-trade-off}
\begin{figure*}[!t]
    \centering

    \subfloat[Geometry]{
        \includegraphics[width=0.48\textwidth]{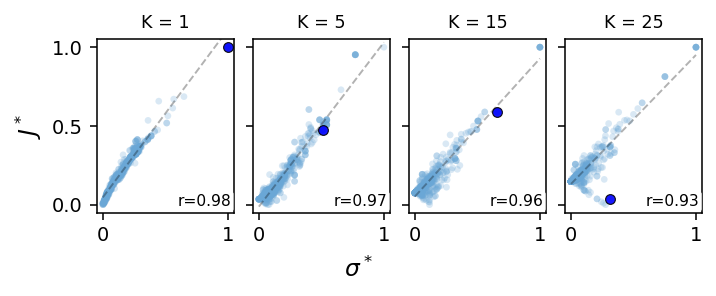}
        \label{fig:1}
    }
    \hfil
    \subfloat[Bars-Rev]{
        \includegraphics[width=0.48\textwidth]{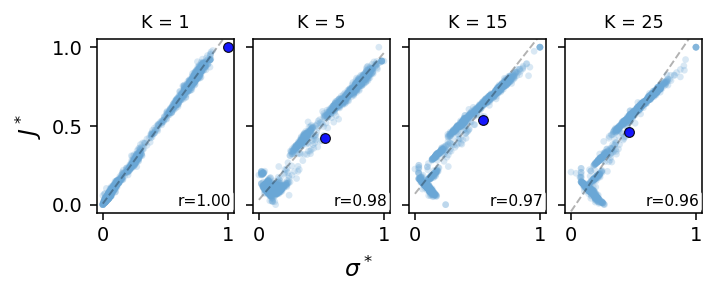}
        \label{fig:2}
    }

    \vspace{0.5em}

    \subfloat[Music-Rev]{
        \includegraphics[width=0.48\textwidth]{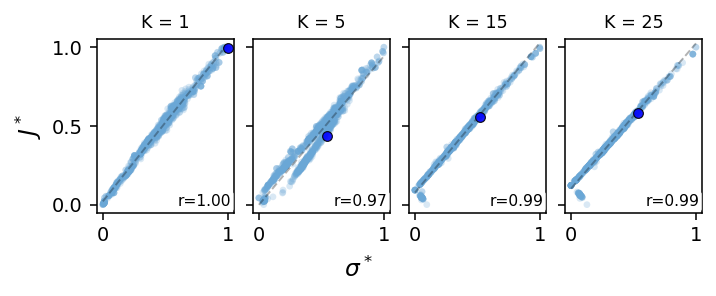}
        \label{fig:3}
    }
    \hfil
    \subfloat[Restaurants-Rev]{
        \includegraphics[width=0.48\textwidth]{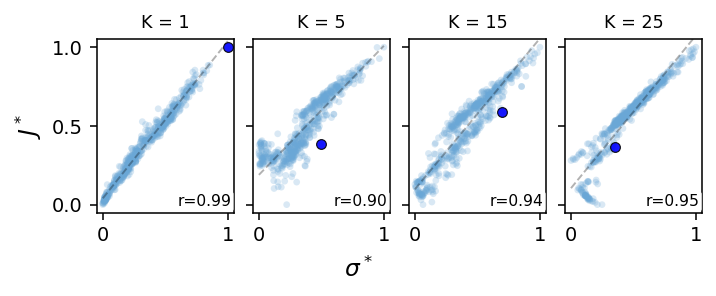}
        \label{fig:4}
    }

    \vspace{0.5em}

    \subfloat[Contact-high-school]{
        \includegraphics[width=0.48\textwidth]{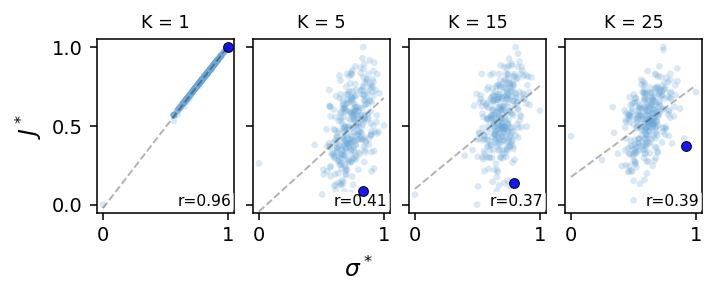}
        \label{fig:5}
    }

    \caption{Relative influence $\sigma^*(u)$ versus relative influence disparity $J^*(u)$ of each candidate node at the FIMH selection step $K \in \{1,5,15,25\}$ when $\beta=0.01$, their Pearson correlation $r$ and linear fit in dashed line in network (a) Gemetry, (b) Bars-Rev, (c) Music-Rev, (d) Restaurants-Rev and (e) Contact-high-school. The blue dot represents the node closest to the utopia point, thus selected as a seed node.}
    \label{fig:correlations}
\end{figure*}

Figure \ref{fig:correlations} shows the relative influence $\sigma^*(u)$ versus relative influence disparity $J^*(u)$ of each candidate node at FIMH selection step $K$ in the other five real-world networks.

\section{Performance evaluation with other process parameters}\label{app:sensitivity-analysis}
The performance of FIMH, FIMH(2-hop) and all the baselines are provided for another two sets of process paramters: $(\beta=0.015, T=15)$ and $(\beta=0.02, T=10)$.

\begin{figure*}[!t]
    \centering
    \includegraphics[width=1\textwidth]{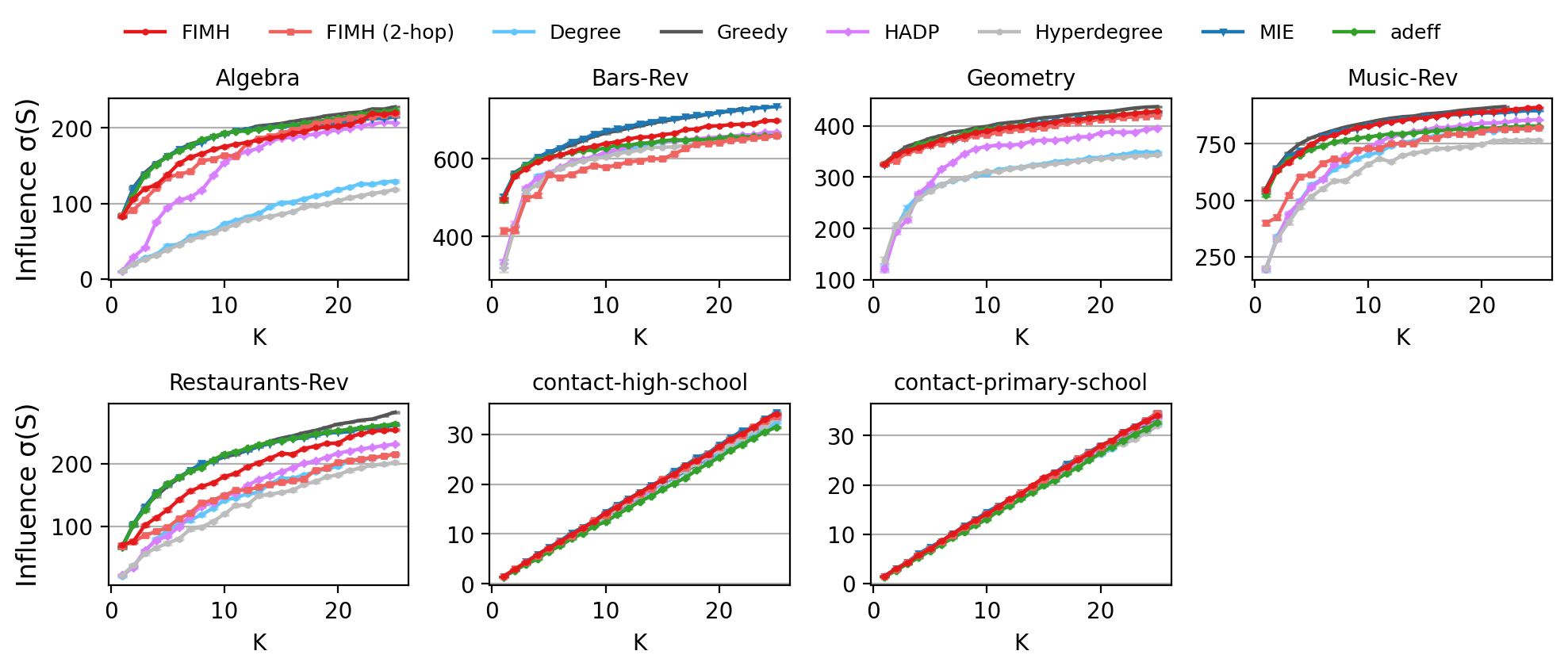}
    \caption{influence $\sigma(S)$ as a function of seed budget $K \in [1,25]$ for all methods under SICP with $\beta=0.015$ and $T=15$.}
    \label{fig:spread1515}
\end{figure*}

\begin{figure*}[!t]
    \centering
    \includegraphics[width=1\textwidth]{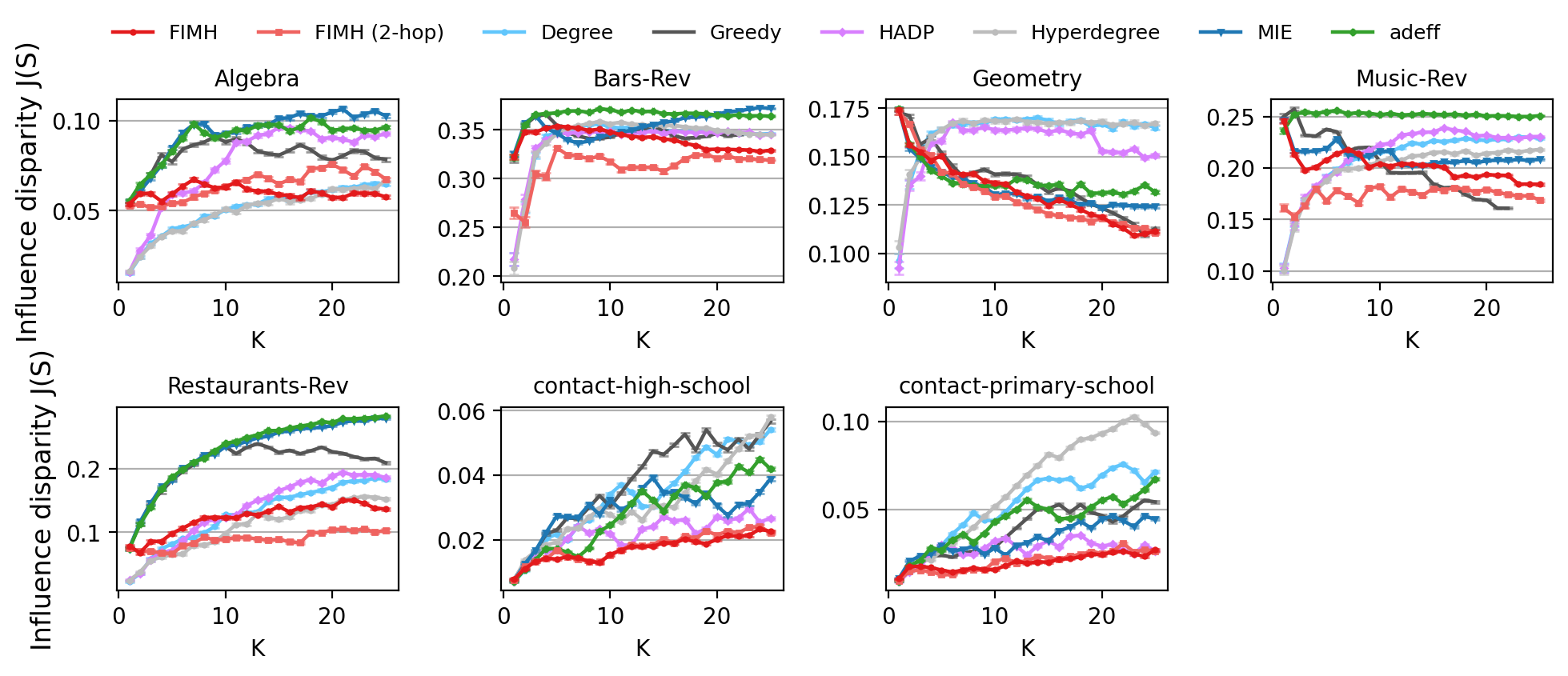}
    \caption{influence disparity $J(S)$ as a function of seed budget $K \in [1,25]$ for all methods under SICP with $\beta=0.015$ and $T=15$.}
    \label{fig:equity1515}
\end{figure*}

\begin{figure*}[!t]
    \centering
    \includegraphics[width=1\textwidth]{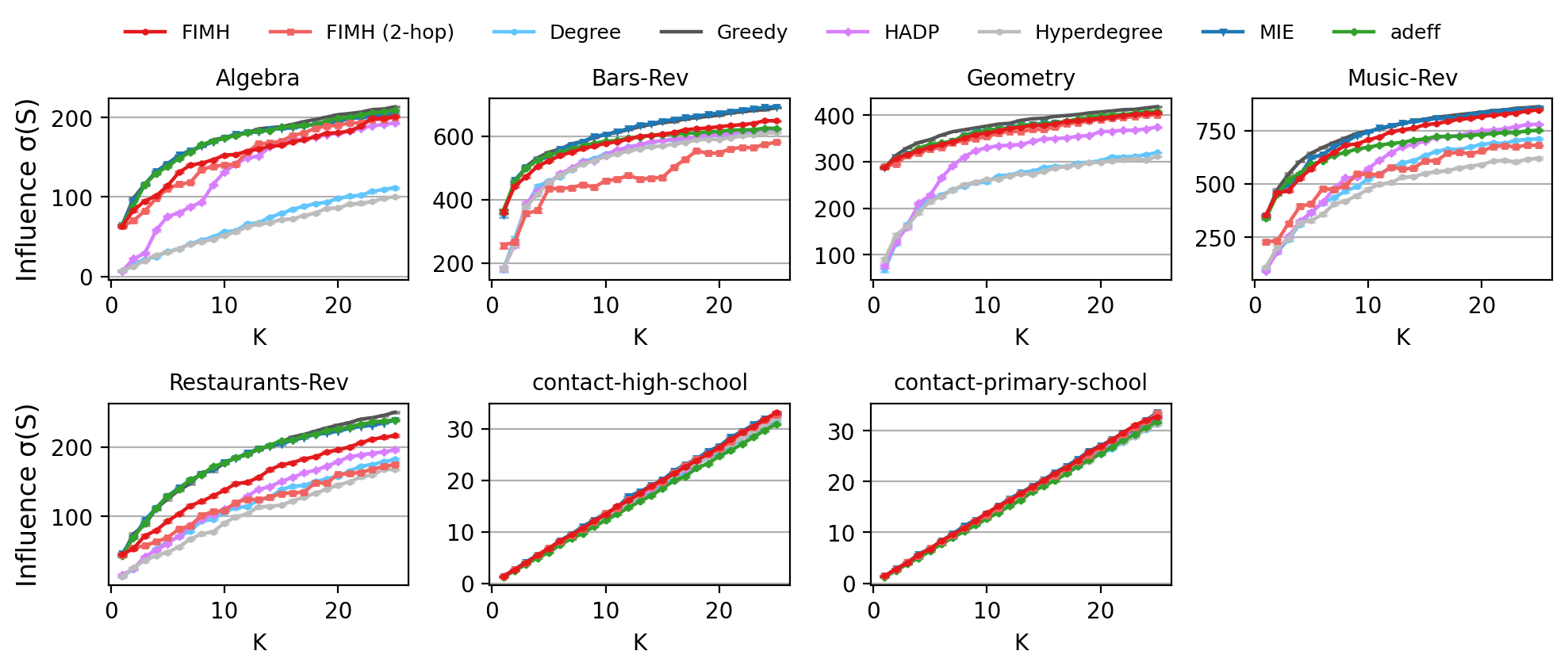}
    \caption{influence $\sigma(S)$ as a function of seed budget $K \in [1,25]$ for all methods under SICP with $\beta=0.02$ and $T=10$.}
    \label{fig:spread2010}
\end{figure*}

\begin{figure*}[!t]
    \centering
    \includegraphics[width=1\textwidth]{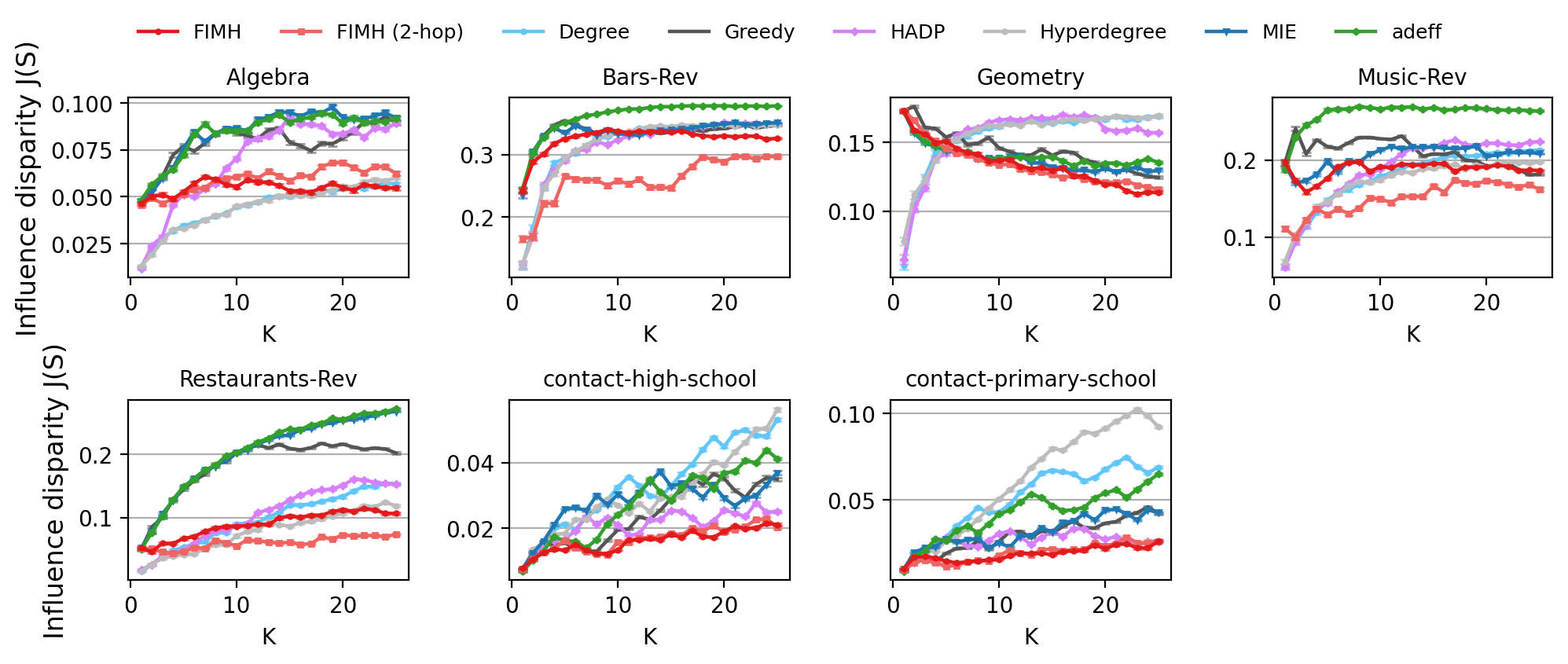}
    \caption{influence disparity $J(S)$ as a function of seed budget $K \in [1,25]$ for all methods under SICP with $\beta=0.02$ and $T=10$.}
    \label{fig:equity2010}
\end{figure*}

Figures \ref{fig:spread1515} and \ref{fig:equity1515} show the influence $\sigma(S)$ and influence disparity $J(S)$, respectively, as a function of the seed budget $K \in [1,25]$ when $\beta = 0.015, T = 15$, respectively. Similarly, Figures \ref{fig:spread2010} and \ref{fig:equity2010} present the performance when $\beta = 0.02, T = 10$. The area under the curve (AUC) for influence and influence disparity resulting respectively from each method are shown in Tables \ref{tab:AUCb015-spread} and \ref{tab:AUCb015-fair} for $\beta = 0.015, T = 15$, and Tables \ref{tab:AUCb020-spread} and \ref{tab:AUCb020-fair} for $(\beta = 0.02, T = 10)$.

\renewcommand{\arraystretch}{1.2}

\begin{table*}[!t]
\centering
\resizebox{\textwidth}{!}{
\begin{tabular}{lrrrrrrr}
\hline
\bfseries Dataset & \bfseries FIMH & \bfseries FIMH (2-hop) & \bfseries MIE & \bfseries Adeff & \bfseries HADP & \bfseries Hdegree & \bfseries Degree \\
\hline
Algebra         & 4429.65    & 4154.60       & 4496.24*  & \textbf{4538.51}  & 3622.62  & 1795.57     & 1987.55    \\
Geometry        & 9446.69*    & 9303.52       & 9431.18   &\textbf{ 9471.09} & 8202.23  & 7304.54     & 7357.21    \\
Bars-Rev        & 15464.00*   & 14088.94      &\textbf{ 16152.26} & 15086.93 & 14634.64 & 14393.77    & 14579.94   \\
Music-Rev       &\textbf{ 19798.97}   & 17238.94      & 19793.40* & 18581.53 & 17244.94 & 15428.50    & 16598.19   \\
Restaurant-Rev & 4524.86    & 3733.54       & 5089.52*   & \textbf{5108.66} & 3799.86  & 3198.52     & 3545.81    \\
High-school     & 432.90*     & 430.25        & \textbf{437.66}   & 395.48   & 420.65   & 421.30      & 413.06     \\
Middle-School   & 436.92*     & 435.52        & \textbf{439.79}   & 410.07   & 428.33   & 414.03      & 414.35     \\
\hline
\end{tabular}
}

\caption{AUC of influence computed for each algorithm curve in Figure \ref{fig:spread1515} when $\beta=0.015, T=15$. The highest AUC in each dataset is shown in bold and the second-highest is marked with an asterisk (*).}
\label{tab:AUCb015-spread}
\end{table*}

\begin{table*}[!t]
\centering
\resizebox{\textwidth}{!}{
\begin{tabular}{lrrrrrrr}
\hline
\bfseries Dataset & \bfseries FIMH & \bfseries FIMH (2-hop) & \bfseries MIE & \bfseries Adeff & \bfseries HADP & \bfseries Hdegree & \bfseries Degree \\
\hline
Algebra         & 1.45    & 1.55            & 2.26     & 2.18   & 1.85  &     \textbf{ 1.20}     & 1.22*   \\
Geometry        & 3.16*    & \textbf{3.10}            & 3.19     & 3.29  &  3.77 &        3.94     & 3.93    \\
Bars-Rev        & 8.18   &\textbf{ 7.53}            & 8.53      & 8.77  & 8.17* &        8.26   & 8.21   \\
Music-Rev       & 4.82*   &\textbf{ 4.18 }           & 5.06       & 6.04 & 5.16 &        4.84    & 5.05   \\
Restaurant-Rev & 2.97    &\textbf{ 2.12}            & 5.54      & 5.61  &  3.29  &      2.61*    & 3.10    \\
High-school     & \textbf{0.41}     & 0.43*          & 0.70      & 0.68   &  0.53  &     0.76      & 0.82     \\
Middle-School   &\textbf{ 0.48}     & 0.49*          & 0.79      & 1.04   &  0.66  &     1.52      & 1.26    \\
\hline
\end{tabular}
}
\caption{AUC of influence disparity computed for each algorithm curve in Figure \ref{fig:equity1515} when $\beta=0.015, T=15$. The lowest AUC in each dataset is shown in bold and the second-lowest is marked with an asterisk (*).}
\label{tab:AUCb015-fair}
\end{table*}

\begin{table*}[!t]
\centering
\resizebox{\textwidth}{!}{
\begin{tabular}{lrrrrrrr}
\hline
\bfseries Dataset & \bfseries FIMH & \bfseries FIMH (2-hop) & \bfseries MIE & \bfseries Adeff & \bfseries HADP & \bfseries Hdegree & \bfseries Degree \\
\hline
Algebra         & 3685.83  & 3646.80      & 4108.85*  & \textbf{4115.43 } & 3194.33    & 1463.93     & 1602.19    \\
Geometry        &  8783.45  & 8654.34       & 8808.08*  &\textbf{ 8861.51} & 7421.40  & 6184.17     & 6244.99    \\
Bars-Rev        &  13886.01* & 11334.53      & \textbf{14668.64} & 13872.50 & 12795.36  & 12565.25    & 12797.29 \\
Music-Rev       &  17043.60*  & 13114.42      & \textbf{17554.34} &15952.19 & 14110.33 & 11481.87    & 12916.40   \\
Restaurant-Rev &  3592.51   & 2853.61       & 4329.53*   &\textbf{4350.30} & 3020.63    & 2453.98     & 2776.51    \\
High-school     &  416.60*     & 415.35        & \textbf{422.31}   & 384.14   & 407.38   & 407.05      & 399.88     \\
Middle-School   & 422.04*     & 418.51       & \textbf{424.41}   & 397.23  & 413.34      & 402.40      &400.83     \\
\hline
\end{tabular}
}
\caption{AUC of influence computed for each algorithm curve in Figure \ref{fig:spread2010} when $\beta=0.02, \; T=10$. The highest AUC in each dataset is shown in bold and the second-highest is marked with an asterisk (*).}
\label{tab:AUCb020-spread}
\end{table*}

\begin{table*}[!t]
\centering
 \resizebox{\textwidth}{!}{
\begin{tabular}{lrrrrrrr}
\hline
\bfseries Dataset & \bfseries FIMH & \bfseries FIMH (2-hop) & \bfseries MIE & \bfseries Adeff & \bfseries HADP & \bfseries Hdegree & \bfseries Degree \\
\hline
Algebra         &  1.32*  & 1.43            & 2.05     & 2.02   & 1.69  &      \textbf{1.08}     & \textbf{1.08 }  \\
Geometry        &   3.21* & \textbf{3.19}            & 3.30     & 3.37  &  3.72 &        3.76     & 3.75    \\
Bars-Rev        & 7.84   & \textbf{6.26}            & 8.09      & 8.73  & 7.61* &        7.66   & 7.68  \\
Music-Rev       & 4.51   & \textbf{3.62}            & 4.91       & 6.28 & 4.56 &       4.15*    & 4.30   \\
Restaurant-Rev & 2.16   &  \textbf{1.46 }         &  4.91      & 4.97  &  2.53  &      1.89*    & 2.35    \\
High-school     & \textbf{0.39}    & 0.40*          & 0.68      & 0.66   &  0.56 &     0.73      & 0.80     \\
Middle-School   & \textbf{0.45}  & 0.46*         & 0.76      & 1.01   &  0.63  &         1.50      & 1.23    \\
\hline
\end{tabular}
}
\caption{AUC of influence disparity computed for each algorithm curve in Figure \ref{fig:equity2010} when $\beta=0.02, T=10$. The lowest AUC in each dataset is shown in bold and the second-lowest is marked with an asterisk (*).}
\label{tab:AUCb020-fair}
\end{table*}

\end{document}